\documentclass[aps,twocolumn,preprintnumbers,amsmath,amssymb,superscriptaddress,floatfix,nofootinbib]{revtex4}
\usepackage{graphicx,color,dcolumn,booktabs,bm}
\usepackage{longtable,lscape}
\usepackage{txfonts}
\usepackage{overpic}
\usepackage{epsfig}
\usepackage{amssymb}
\usepackage{rotating}
\usepackage{epstopdf}
\usepackage{indentfirst}
\usepackage{feynmf}   %{feynmp}
\usepackage{slashed}  %for Feynman symbols
\usepackage{cases}
\usepackage{color}
\usepackage{float}
\usepackage{multirow}
\usepackage{graphicx,color,dcolumn,booktabs,bm}
\usepackage{cases}
\usepackage{array}

\graphicspath{{Figures/}} %

\usepackage[colorlinks, citecolor=blue,anchorcolor=red,menucolor=red,
linkcolor=red,filecolor=red,runcolor=red,urlcolor=blue,frenchlinks=red]{hyperref}

\begin{document}

\title{Open charm and bottom tetraquarks in an extended relativized quark model}

\author{Qi-Fang L\"u} \email{lvqifang@hunnu.edu.cn}
\affiliation{  Department
of Physics, Hunan Normal University,  Changsha 410081, China }

\affiliation{ Synergetic Innovation
Center for Quantum Effects and Applications (SICQEA), Changsha 410081,China}

\affiliation{  Key Laboratory of
Low-Dimensional Quantum Structures and Quantum Control of Ministry
of Education, Changsha 410081, China}
\author{Dian-Yong Chen} \email{chendy@seu.edu.cn}
\affiliation{School of Physics, Southeast University, Nanjing 210094, China }
\author{Yu-Bing Dong} \email{dongyb@ihep.ac.cn}
\affiliation{Institute of High Energy Physics, Chinese Academy of Sciences, Beijing 100049, China}
\affiliation{Theoretical Physics Center for Science Facilities (TPCSF), CAS, Beijing 100049, China}
\affiliation{School of Physical Sciences, University of Chinese Academy of Sciences, Beijing 101408, China}

\begin{abstract}
In present work, we systematically calculate the mass spectra of open charm and bottom tetraquarks $q q\bar q \bar Q$ within an extended relativized quark model. The four-body relativized Hamiltonians including the Coulomb potential, confining potential, spin-spin interactions, and relativistic corrections are solved by using the variational method. It is found that the predicted masses of four $0^+$ $ud\bar s \bar c$ states are 2765, 3065, 3152, and 3396 MeV, which disfavor the assignment of the newly observed $X_0(2900)$ as a compact tetraquark. Moreover, the whole mass spectra of the open charm and bottom tetraquarks show quite similar patterns, which preserve the light flavor SU(3) symmetry and heavy quark symmetry well.
In addition, our results suggest that the flavor exotic states $nn\bar s \bar c$, $nn\bar s \bar b$, $ss\bar n \bar c$, and
$ss\bar n \bar b$ and their antiparticles can be searched in the heavy-light meson plus kaon final states by future experiments. More theoretical and experimental efforts are needed to investigate these singly heavy tetraquarks.
\end{abstract}

\maketitle

\section{Introduction}{\label{Sec:Int}}

In the past years, many new hadronic states have been observed experimentally, and some of them cannot be simply assigned into the
conventional mesons or baryons. This significant progress in experiments has triggered plenty of theoretical interests and made the
study of those exotic states as one of the intriguing topics in hadronic physics~\cite{Klempt:2007cp,Brambilla:2010cs,Hosaka:2016pey,Chen:2016qju,Lebed:2016hpi,Esposito:2016noz,Dong:2017gaw,Ali:2017jda,Guo:2017jvc,Olsen:2017bmm,Karliner:2017qhf,Liu:2019zoy,Brambilla:2019esw,Richard:2019cmi}. Among those states, the charged resonances $Z_{c(b)}$~\cite{Choi:2007wga,Aaij:2014jqa,Belle:2011aa,Ablikim:2013mio,Liu:2013dau},
fully heavy tetraquark X(6900)~\cite{Aaij:2020fnh}, and pentaquarks $P_c$~\cite{Aaij:2015tga,Aaij:2019vzc}, are particularly interesting, since they cannot mix with traditional hadrons
in the heavy quark sectors. Besides hidden charm and bottom states, the existence of flavor exotic states, where quarks and
antiquarks cannot annihilate though strong and electromagnetic interactions, was also predicted. Therefore, searching for these
flavor exotic states have become increasingly important both theoretically and experimentally.

In 2016, the D0 Collaboration reported the evidence of a narrow structure $X(5568)$, which is expected to be composed of four different
flavors~\cite{D0:2016mwd}. Unfortunately, its existence was not confirmed by the LHCb, CMS, CDF, and ATLAS Collaborations~\cite{Aaij:2016iev,Sirunyan:2017ofq,Aaltonen:2017voc,Aaboud:2018hgx}, though the D0 Collaboration claimed that the $X(5568)$ was also found in different decay chains~\cite{Abazov:2017poh}. Before the observation of $X(5568)$, there existed a few studies
on open charm and bottom tetraquark states, which mainly concentrated on the tetraquark interpretation of $D_{s0}^*(2317)$~\cite{Bracco:2005kt,Maiani:2004vq,Vijande:2006hj,Carlucci:2007um,Zhang:2006hv,Gerasyuta:2008ps,Ebert:2010af}. The evidence of $X(5568)$ immediately attracted great interests and many extensive theoretical investigations under various
interpretations, such as tetraquark~\cite{Agaev:2016mjb,Wang:2016tsi,Wang:2016mee,Zanetti:2016wjn,Chen:2016mqt,Agaev:2016ijz,Liu:2016ogz,Dias:2016dme,Wang:2016wkj,Stancu:2016sfd,Tang:2016pcf,Ali:2016gdg,Agaev:2016srl,Goerke:2016hxf,Agaev:2016ifn,Agamaliev:2016wtt,Lu:2016zhe,Chen:2017rhl,Zhang:2017xwc}, molecule~\cite{Xiao:2016mho,Agaev:2016urs,Albaladejo:2016eps,Chen:2016npt,Kang:2016zmv,Lang:2016jpk,Chen:2016ypj,Lu:2016kxm,Sun:2016tmz,Ke:2016oez}, and kinematic effects~\cite{Liu:2016xly,Yang:2016sws,Liu:2017vsf}. Other related topics are also wildly discussed~\cite{Esposito:2016itg,Agaev:2016lkl,He:2016yhd,Jin:2016zuy,Burns:2016gvy,Guo:2016nhb,He:2016xvd,Yu:2017pmn,Azizi:2018mte,Chen:2018hts,Huang:2019otd,Xing:2019hjg,Agaev:2019wkk,Cheng:2020nho}, and the review on $X(5568)$ can be found in Ref.~\cite{Chen:2016qju}. Basically, the experimental and theoretical efforts indicated that the $X(5568)$ should not be a genuine resonance. Although the searches of $X(5568)$ failed, the investigations of open charm and bottom tetraquark states have been revitalized.

Very recently, the LHCb Collaboration reported the observation of an exotic structure near $2.9 ~\rm{GeV}$ in the $D^-K^+$
invariant mass spectrum via the $B^+ \to D^+ D^- K^+$ decay channel~\cite{lhc}. Then, this peak is modelled according to two
resonances, $X_0(2900)$ and $X_1(2900)$. Their parameters are fitted to be
\begin{eqnarray}
	m[X_0(2900)] &=& 2866.3 \pm 6.5 \pm 2.0 ~\mathrm{MeV}, \nonumber \\
    \Gamma[X_0(2900)]	&=& 57.2 \pm 12.2 \pm 4.1 ~\mathrm{MeV},
\end{eqnarray}
\begin{eqnarray}
	m[X_1(2900)] &=& 2904.1 \pm 4.8 \pm 1.3 ~\mathrm{MeV}, \nonumber \\
    \Gamma[X_1(2900)]	&=& 110.3 \pm 10.7 \pm 4.3 ~\mathrm{MeV}.
\end{eqnarray}
Given their $D^-K^+$ decay mode, the quantum numbers of $X_0(2900)$ and $X_1(2900)$ should be $J^P=0^+$ and $1^-$, respectively.
Also, both of them have four different flavors, which indicates their exotic nature.

After the observation of the LHCb Collaboration, these two states near $2.9~\rm{GeV}$ were discussed within the simple quark model~\cite{Karliner:2020vsi}.
The $X_0(2900)$ was interpreted as an isosinglet compact tetraquark state,
while the $X_1(2900)$ may be regarded as an artifact due to rescattering effects or a $J^P=2^+$ $\bar D^*K^*$ molecule~\cite{Karliner:2020vsi}. Until now, no rigorous four-body calculation for these two states
does exist, and all sorts of explanations are possible. Therefore, it is essential to investigate the possible compact tetraquark interpretations of $X_0(2900)$ and
$X_1(2900)$ within realistic potentials.

In Refs.~\cite{Lu:2020rog,Lu:2020cns}, we have extended the relativized quark model proposed by Godfrey and Isgur to investigate
the doubly and fully heavy tetraquarks with the original model parameters. This extension allows us to describe the tetraquarks 
and conventional mesons in a uniform frame. Since the relativized potential can give a unified description of different flavor
sectors and involve relativistic effects, it is believed to be more suitable to deal with the heavy-light and light-light quark
interactions. In present work, we will systematically investigate the open charm and bottom tetraquarks $qq\bar{q}\bar{Q}$ in the extended
relativized quark model and test the possible assignments of the newly observed resonances.

This paper is organized as follows. In Section~\ref{model}, we briefly introduce the formalism of our extended relativized quark
model for tetraquarks. Then, the mass spectra and discussions of our numerical results are
presented in Section~\ref{results}. Finally, a short summary is given in the last section.   

\section{Extended relativized quark model}{\label{model}}

To investigate the $S-$wave mass spectra of open charm and bottom tetraquarks $q_1 q_2 \bar q_3 \bar Q_4$, the extended relativized
quark model is employed~\cite{Lu:2020rog}. This model is a natural generalization of the relativized quark model to deal with the
tetraquark states. The relevant Hamiltonian with quark and gluon degrees of freedom for a $q_1 q_2 \bar q_3 \bar Q_4$ state can be written
as
\begin{equation}
H = H_0+\sum_{i<j}V_{ij}^{\rm oge}+\sum_{i<j}V_{ij}^{\rm conf}, \label{ham}
\end{equation}
where
\begin{equation}
H_0 = \sum_{i=1}^{4}(p_i^2+m_i^2)^{1/2}
\end{equation}
is the relativistic kinetic energy, $V_{ij}^{\rm oge}$ is the one gluon exchange potential together with the spin-spin interactions,
and $V_{ij}^{\rm conf}$ corresponds to the confinement potential. The explicit formula and parameters of these relativized potentials
can be found in Refs.~\cite{Godfrey:1985xj,Lu:2020rog}.

The wave function for a $q_1 q_2 \bar q_3 \bar Q_4$ state is constituted of four different parts: color, flavor, spin, and spatial
wave functions.
For the color part, there are two types of colorless states with certain permutation properties,
\begin{equation}
|\bar 3 3\rangle = |(q_1 q_2)^{\bar 3} (\bar q_3 \bar Q_4)^3\rangle,
\end{equation}
\begin{equation}
|6 \bar 6\rangle = |(q_1 q_2)^{6} (\bar q_3 \bar Q_4)^{\bar 6}\rangle.
\end{equation}
Here, the $|\bar 3 3\rangle$ and $|6 \bar 6 \rangle$ are antisymmetric and symmetric color wave functions under the exchange of $q_1q_2$ or $\bar q_3 \bar Q_4$, respectively. For the flavor part, the combination $q_1q_2$ can be either symmetric or antisymmetric, while the $\bar q_3$ and $\bar Q_4$ are treated as different particles without symmetry constraint. To distinguish the up, down, and strange quarks clearly, we adopt the notation "$n$" to stand for the up or down quark, and "$s$" to represent the strange quark.

In the spin space, the six spin bases can be expressed as
\begin{equation}
\chi^{00}_0 = |(q_1 q_2)_0 (\bar q_3 \bar Q_4)_0\rangle_0,
\end{equation}
\begin{equation}
\chi^{11}_0 = |(q_1 q_2)_1 (\bar q_3 \bar Q_4)_1\rangle_0,
\end{equation}
\begin{equation}
\chi^{01}_1 = |(q_1 q_2)_0 (\bar q_3 \bar Q_4)_1\rangle_1,
\end{equation}
\begin{equation}
\chi^{10}_1 = |(q_1 q_2)_1 (\bar q_3 \bar Q_4)_0\rangle_1,
\end{equation}
\begin{equation}
\chi^{11}_1 = |(q_1 q_2)_1 (\bar q_3 \bar Q_4)_1\rangle_1,
\end{equation}
\begin{equation}
\chi^{11}_2 = |(q_1 q_2)_1 (\bar q_3 \bar Q_4)_1\rangle_2,
\end{equation}
where $(q_1 q_2)_0$ and $(\bar q_3 \bar Q_4)_0$ are antisymmetric for the two fermions under permutations, while the
$(q_1 q_2)_1$ and $(\bar q_3 \bar Q_4)_1$ are symmetric ones. The relevant matrix elements of the color and spin parts for various types of tetraquark states are identical~\cite{Lu:2020rog}.

\begin{figure*}[!htbp]
\includegraphics[scale=0.7]{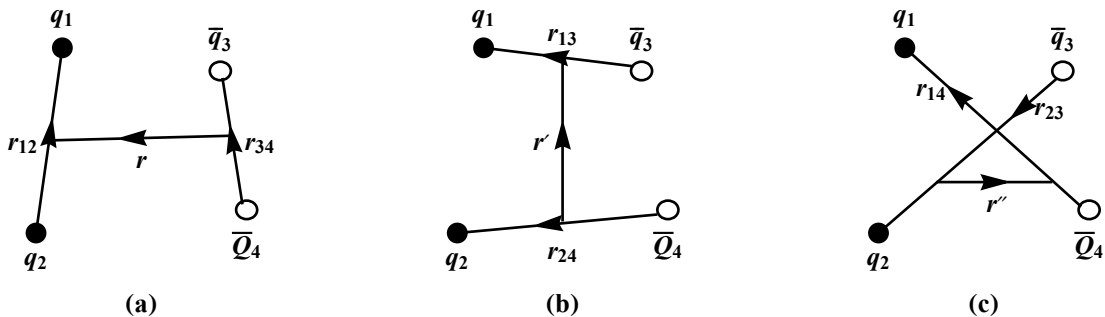}
\vspace{0.0cm} \caption{The $q_1q_2 \bar q_3 \bar Q_4$ tetraquark state in Jacobi coordinates.}
\label{jacobi}
\end{figure*}

In the spatial space, the Jacobi coordinates are shown in Fig.~\ref{jacobi}. For a $q_1 q_2 \bar q_3 \bar Q_4$ state,
one can define
\begin{equation}
\boldsymbol r_{12}=  \boldsymbol r_1 - \boldsymbol r_2,
\end{equation}
\begin{equation}
\boldsymbol r_{34}=  \boldsymbol r_3 - \boldsymbol r_4,
\end{equation}
\begin{equation}
\boldsymbol r = \frac{m_1 \boldsymbol r_1 + m_2 \boldsymbol r_2}{m_1+m_2} - \frac{m_3 \boldsymbol r_3 + m_4 \boldsymbol r_4}{m_3+m_4},
\end{equation}
and
\begin{equation}
\boldsymbol R = \frac{m_1 \boldsymbol r_1 + m_2 \boldsymbol r_2 + m_3 \boldsymbol r_3 + m_4 \boldsymbol r_4}{m_1+m_2+m_3+m_4}.
\end{equation}
Then, other coordinates of this system can be expressed in terms of $\boldsymbol r_{12}$, $\boldsymbol r_{34}$,
and $\boldsymbol r$~\cite{Lu:2020rog}. A set of Gaussian
functions is adopted to approach the $S-$wave realistic spatial wave function~\cite{Hiyama:2003cu}
\begin{equation}
\Psi(\boldsymbol r_{12},\boldsymbol r_{34},\boldsymbol r) = \sum_{n_{12},n_{34},n} C_{n_{12}n_{34}n}
\psi_{n_{12}}(\boldsymbol r_{12}) \psi_{n_{34}}(\boldsymbol r_{34}) \psi_n(\boldsymbol r),
\end{equation}
where $C_{n_{12}n_{34}n}$ are the expansion coefficients. The $\psi_{n_{12}}(\boldsymbol r_{12}) \psi_{n_{34}}(\boldsymbol r_{34})
\psi_n(\boldsymbol r)$ is the position representation of the basis $|n_{12}n_{34}n\rangle$, where
\begin{equation}
\psi_n(\boldsymbol r) = \frac{2^{7/4}\nu_n^{3/4}}{\pi^{1/4}} e^{-\nu_n r^2} Y_{00}(\hat{\boldsymbol r}) =
\Bigg(\frac{2 \nu_n}{\pi} \Bigg )^{3/4} e^{-\nu_n r^2},
\end{equation}
\begin{equation}
\nu_n = \frac{1}{r_1^2a^{2(n-1)}},~~~~ (n=1-N_{max}).
\end{equation}
The final results are independent with geometric Gaussian size parameters $r_1$, $a$, and $N_{max}$ when
adequate bases are chosen~\cite{Hiyama:2003cu}. The $\psi_{n_{12}}(\boldsymbol r_{12})$ and  
$\psi_{n_{34}}(\boldsymbol r_{34})$ can be expressed in a similar way, and the momentum representation of the 
basis $ |n_{12}n_{34}n\rangle$ can be obtained via Fourier transformation. The numerical error of our approach 
has been analyzed in Ref.~\cite{Lu:2020rog}, which is sufficient for quark model predictions. 

According to the Pauli exclusion principle, the wave function of a tetraquark should be antisymmetric for the identical quarks and antiquarks. All possible configurations for the
$q_1q_2 \bar q_3 \bar Q_4$ systems are listed in Table~\ref{configuration}. For the notations, the subscripts and superscripts are the spin and color types, respectively. The brackets $[~]$ and braces $\{ ~\}$ stand for the antisymmetric and symmetric flavor wave functions, respectively. The parentheses $(~)$  are adopted for the subsystems without permutation symmetries. 

With the total wave functions, all the matrix elements involved in the Hamiltonian can be worked out straightforwardly. The masses without mixing mechanism can be obtained by solving the generalized eigenvalue problem
\begin{equation}
\sum_{j=1}^{N_{max}^3}(H_{ij}-EN_{ij})C_j=0,~~~~ (i=1-N_{max}^3),
\end{equation}
where the $H_{ij}$ are the matrix elements of Hamiltonian, $N_{ij}$ are the overlap matrix elements of the Gaussian bases due to their
nonorthogonality, $E$ is the mass, and $C_j$
is the eigenvector corresponding to the expansion coefficients $C_{n_{12}n_{34}n}$ for the spatial wave function. For a given system, different configurations with same $I(J^{P})$ should mix with each other.  The final mass spectra and wave functions are obtained by diagonalizing the mass matrix of these configurations.

\begin{table}[!htbp]
\begin{center}
\footnotesize
\caption{ \label{configuration} All possible configurations for the $q_1q_2\bar q_3 \bar Q_4$ systems.}
\begin{tabular*}{8.7cm}{@{\extracolsep{\fill}}p{1cm}<{\centering}p{1.0cm}<{\centering}p{1.7cm}<{\centering}p{1.7cm}<{\centering}p{1.7cm}<{\centering}}
\hline\hline
System & $I(J^P)$         &  \multicolumn{3}{c}{Configuration} \\\hline

$[nn] (\bar n \bar c)$ & $\frac{1}{2}(0^+)$ &  $|[nn]^{\bar 3}_0 (\bar n \bar c)^3_0 \rangle_0$ &  $|[nn]^6_1 (\bar n \bar c)^{\bar 6}_1\rangle_0$     & $\cdots$\\
& $\frac{1}{2}(1^+)$       &  $|[nn]^{\bar 3}_0 (\bar n \bar c)^3_1\rangle_1$     &  $|[nn]^6_1 (\bar n \bar c)^{\bar 6}_0\rangle_1$   &  $|[nn]^6_1 (\bar n \bar c)^{\bar 6}_1\rangle_1$    \\
& $\frac{1}{2}(2^+)$       &   $|[nn]^6_1 (\bar n \bar c)^{\bar 6}_1\rangle_2$   & $\cdots$  & $\cdots$    \\
$\{nn\} (\bar n \bar c)$& $\frac{1}{2}/ \frac{3}{2}(0^+)$       & $|\{nn\}^{\bar 3}_1 (\bar n \bar c)^3_1\rangle_0$     &  $|\{nn\}^6_0 (\bar n \bar c)^{\bar 6}_0\rangle_0$   & $\cdots$    \\
& $\frac{1}{2}/ \frac{3}{2}(1^+)$      &  $|\{nn\}^{\bar 3}_1 (\bar n \bar c)^3_0\rangle_1$  &  $|\{nn\}^{\bar 3}_1 (\bar n \bar c)^3_1\rangle_1$     &   $|\{nn\}^6_0 (\bar n \bar c)^{\bar 6}_1\rangle_1$     \\
& $\frac{1}{2} / \frac{3}{2}(2^+)$       &  $|\{nn\}^{\bar 3}_1 (\bar n \bar c)^3_1\rangle_2$     &  $\cdots$   & $\cdots$     \\

$[nn] (\bar n \bar b)$ & $\frac{1}{2}(0^+)$ &  $|[nn]^{\bar 3}_0 (\bar n \bar b)^3_0 \rangle_0$ &  $|[nn]^6_1 (\bar n \bar b)^{\bar 6}_1\rangle_0$     & $\cdots$\\
& $\frac{1}{2}(1^+)$       &  $|[nn]^{\bar 3}_0 (\bar n \bar b)^3_1\rangle_1$     &  $|[nn]^6_1 (\bar n \bar b)^{\bar 6}_0\rangle_1$   &  $|[nn]^6_1 (\bar n \bar b)^{\bar 6}_1\rangle_1$    \\
& $\frac{1}{2}(2^+)$       &   $|[nn]^6_1 (\bar n \bar b)^{\bar 6}_1\rangle_2$   & $\cdots$  & $\cdots$    \\
$\{nn\} (\bar n \bar b)$& $\frac{1}{2}/ \frac{3}{2}(0^+)$       & $|\{nn\}^{\bar 3}_1 (\bar n \bar b)^3_1\rangle_0$     &  $|\{nn\}^6_0 (\bar n \bar b)^{\bar 6}_0\rangle_0$   & $\cdots$    \\
& $\frac{1}{2}/ \frac{3}{2}(1^+)$      &  $|\{nn\}^{\bar 3}_1 (\bar n \bar b)^3_0\rangle_1$  &  $|\{nn\}^{\bar 3}_1 (\bar n \bar b)^3_1\rangle_1$     &   $|\{nn\}^6_0 (\bar n \bar b)^{\bar 6}_1\rangle_1$     \\
& $\frac{1}{2}/ \frac{3}{2}(2^+)$       &  $|\{nn\}^{\bar 3}_1 (\bar n \bar b)^3_1\rangle_2$     &  $\cdots$   & $\cdots$     \\\hline

$[nn] (\bar s \bar c)$ & $0(0^+)$ &  $|[nn]^{\bar 3}_0 (\bar s \bar c)^3_0 \rangle_0$ &  $|[nn]^6_1 (\bar s \bar c)^{\bar 6}_1\rangle_0$     & $\cdots$\\
& $0(1^+)$       &  $|[nn]^{\bar 3}_0 (\bar s \bar c)^3_1\rangle_1$     &  $|[nn]^6_1 (\bar s \bar c)^{\bar 6}_0\rangle_1$   &  $|[nn]^6_1 (\bar s \bar c)^{\bar 6}_1\rangle_1$    \\
& $0(2^+)$       &   $|[nn]^6_1 (\bar s \bar c)^{\bar 6}_1\rangle_2$   & $\cdots$  & $\cdots$    \\
$\{nn\} (\bar s \bar c)$& $1(0^+)$       & $|\{nn\}^{\bar 3}_1 (\bar s \bar c)^3_1\rangle_0$     &  $|\{nn\}^6_0 (\bar s \bar c)^{\bar 6}_0\rangle_0$   & $\cdots$    \\
& $1(1^+)$      &  $|\{nn\}^{\bar 3}_1 (\bar s \bar c)^3_0\rangle_1$  &  $|\{nn\}^{\bar 3}_1 (\bar s \bar c)^3_1\rangle_1$     &   $|\{nn\}^6_0 (\bar s \bar c)^{\bar 6}_1\rangle_1$     \\
& $1(2^+)$       &  $|\{nn\}^{\bar 3}_1 (\bar s \bar c)^3_1\rangle_2$     &  $\cdots$   & $\cdots$     \\

$[nn] (\bar s \bar b)$ & $0(0^+)$ &  $|[nn]^{\bar 3}_0 (\bar s \bar b)^3_0 \rangle_0$ &  $|[nn]^6_1 (\bar s \bar b)^{\bar 6}_1\rangle_0$     & $\cdots$\\
& $0(1^+)$       &  $|[nn]^{\bar 3}_0 (\bar s \bar b)^3_1\rangle_1$     &  $|[nn]^6_1 (\bar s \bar b)^{\bar 6}_0\rangle_1$   &  $|[nn]^6_1 (\bar s \bar b)^{\bar 6}_1\rangle_1$    \\
& $0(2^+)$       &   $|[nn]^6_1 (\bar s \bar b)^{\bar 6}_1\rangle_2$   & $\cdots$  & $\cdots$    \\
$\{nn\} (\bar s \bar b)$& $1(0^+)$       & $|\{nn\}^{\bar 3}_1 (\bar s \bar b)^3_1\rangle_0$     &  $|\{nn\}^6_0 (\bar s \bar b)^{\bar 6}_0\rangle_0$   & $\cdots$    \\
& $1(1^+)$      &  $|\{nn\}^{\bar 3}_1 (\bar s \bar b)^3_0\rangle_1$  &  $|\{nn\}^{\bar 3}_1 (\bar s \bar b)^3_1\rangle_1$     &   $|\{nn\}^6_0 (\bar s \bar b)^{\bar 6}_1\rangle_1$     \\
& $1(2^+)$       &  $|\{nn\}^{\bar 3}_1 (\bar s \bar b)^3_1\rangle_2$     &  $\cdots$   & $\cdots$     \\\hline

$[ns] (\bar n \bar c)$ & $0/1(0^+)$ &  $|[ns]^{\bar 3}_0 (\bar n \bar c)^3_0 \rangle_0$ &  $|[ns]^6_1 (\bar n \bar c)^{\bar 6}_1\rangle_0$     & $\cdots$\\
& $0/1(1^+)$       &  $|[ns]^{\bar 3}_0 (\bar n \bar c)^3_1\rangle_1$     &  $|[ns]^6_1 (\bar n \bar c)^{\bar 6}_0\rangle_1$   &  $|[ns]^6_1 (\bar n \bar c)^{\bar 6}_1\rangle_1$    \\
& $0/1(2^+)$       &   $|[ns]^6_1 (\bar n \bar c)^{\bar 6}_1\rangle_2$   & $\cdots$  & $\cdots$    \\
$\{ns\} (\bar n \bar c)$& $0/1(0^+)$       & $|\{ns\}^{\bar 3}_1 (\bar n \bar c)^3_1\rangle_0$     &  $|\{ns\}^6_0 (\bar n \bar c)^{\bar 6}_0\rangle_0$   & $\cdots$    \\
& $0/1(1^+)$      &  $|\{ns\}^{\bar 3}_1 (\bar n \bar c)^3_0\rangle_1$  &  $|\{ns\}^{\bar 3}_1 (\bar n \bar c)^3_1\rangle_1$     &   $|\{ns\}^6_0 (\bar n \bar c)^{\bar 6}_1\rangle_1$     \\
& $0/1(2^+)$       &  $|\{ns\}^{\bar 3}_1 (\bar n \bar c)^3_1\rangle_2$     &  $\cdots$   & $\cdots$     \\

$[ns] (\bar n \bar b)$ & $0/1(0^+)$ &  $|[ns]^{\bar 3}_0 (\bar n \bar b)^3_0 \rangle_0$ &  $|[ns]^6_1 (\bar n \bar b)^{\bar 6}_1\rangle_0$     & $\cdots$\\
& $0/1(1^+)$       &  $|[ns]^{\bar 3}_0 (\bar n \bar b)^3_1\rangle_1$     &  $|[ns]^6_1 (\bar n \bar b)^{\bar 6}_0\rangle_1$   &  $|[ns]^6_1 (\bar n \bar b)^{\bar 6}_1\rangle_1$    \\
& $0/1(2^+)$       &   $|[ns]^6_1 (\bar n \bar b)^{\bar 6}_1\rangle_2$   & $\cdots$  & $\cdots$    \\
$\{ns\} (\bar n \bar b)$& $0/1(0^+)$       & $|\{ns\}^{\bar 3}_1 (\bar n \bar b)^3_1\rangle_0$     &  $|\{ns\}^6_0 (\bar n \bar b)^{\bar 6}_0\rangle_0$   & $\cdots$    \\
& $0/1(1^+)$      &  $|\{ns\}^{\bar 3}_1 (\bar n \bar b)^3_0\rangle_1$  &  $|\{ns\}^{\bar 3}_1 (\bar n \bar b)^3_1\rangle_1$     &   $|\{ns\}^6_0 (\bar n \bar b)^{\bar 6}_1\rangle_1$     \\
& $0/1(2^+)$       &  $|\{ns\}^{\bar 3}_1 (\bar n \bar b)^3_1\rangle_2$     &  $\cdots$   & $\cdots$     \\\hline

$[ns] (\bar s \bar c)$ & $\frac{1}{2}(0^+)$ &  $|[ns]^{\bar 3}_0 (\bar s \bar c)^3_0 \rangle_0$ &  $|[ns]^6_1 (\bar s \bar c)^{\bar 6}_1\rangle_0$     & $\cdots$\\
& $\frac{1}{2}(1^+)$       &  $|[ns]^{\bar 3}_0 (\bar s \bar c)^3_1\rangle_1$     &  $|[ns]^6_1 (\bar s \bar c)^{\bar 6}_0\rangle_1$   &  $|[ns]^6_1 (\bar s \bar c)^{\bar 6}_1\rangle_1$    \\
& $\frac{1}{2}(2^+)$       &   $|[ns]^6_1 (\bar s \bar c)^{\bar 6}_1\rangle_2$   & $\cdots$  & $\cdots$    \\
$\{ns\} (\bar s \bar c)$& $\frac{1}{2}(0^+)$       & $|\{ns\}^{\bar 3}_1 (\bar s \bar c)^3_1\rangle_0$     &  $|\{ns\}^6_0 (\bar s \bar c)^{\bar 6}_0\rangle_0$   & $\cdots$    \\
& $\frac{1}{2}(1^+)$      &  $|\{ns\}^{\bar 3}_1 (\bar s \bar c)^3_0\rangle_1$  &  $|\{ns\}^{\bar 3}_1 (\bar s \bar c)^3_1\rangle_1$     &   $|\{ns\}^6_0 (\bar s \bar c)^{\bar 6}_1\rangle_1$     \\
& $\frac{1}{2}(2^+)$       &  $|\{ns\}^{\bar 3}_1 (\bar s \bar c)^3_1\rangle_2$     &  $\cdots$   & $\cdots$     \\

$[ns] (\bar s \bar b)$ & $\frac{1}{2}(0^+)$ &  $|[ns]^{\bar 3}_0 (\bar s \bar b)^3_0 \rangle_0$ &  $|[ns]^6_1 (\bar s \bar b)^{\bar 6}_1\rangle_0$     & $\cdots$\\
& $\frac{1}{2}(1^+)$       &  $|[ns]^{\bar 3}_0 (\bar s \bar b)^3_1\rangle_1$     &  $|[ns]^6_1 (\bar s \bar b)^{\bar 6}_0\rangle_1$   &  $|[ns]^6_1 (\bar s \bar b)^{\bar 6}_1\rangle_1$    \\
& $\frac{1}{2}(2^+)$       &   $|[ns]^6_1 (\bar s \bar b)^{\bar 6}_1\rangle_2$   & $\cdots$  & $\cdots$    \\
$\{ns\} (\bar s \bar b)$& $\frac{1}{2}(0^+)$       & $|\{ns\}^{\bar 3}_1 (\bar s \bar b)^3_1\rangle_0$     &  $|\{ns\}^6_0 (\bar s \bar b)^{\bar 6}_0\rangle_0$   & $\cdots$    \\
& $\frac{1}{2}(1^+)$      &  $|\{ns\}^{\bar 3}_1 (\bar s \bar b)^3_0\rangle_1$  &  $|\{ns\}^{\bar 3}_1 (\bar s \bar b)^3_1\rangle_1$     &   $|\{ns\}^6_0 (\bar s \bar b)^{\bar 6}_1\rangle_1$     \\
& $\frac{1}{2}(2^+)$       &  $|\{ns\}^{\bar 3}_1 (\bar s \bar b)^3_1\rangle_2$     &  $\cdots$   & $\cdots$     \\\hline

$\{ss\}(\bar n \bar c)$ & $\frac{1}{2}(0^+)$       & $|\{ss\}^{\bar 3}_1 (\bar n \bar c)^3_1\rangle_0$     &  $|\{ss\}^6_0 (\bar n \bar c)^{\bar 6}_0\rangle_0$   & $\cdots$     \\
& $\frac{1}{2}(1^+)$      &  $|\{ss\}^{\bar 3}_1 (\bar n \bar c)^3_0\rangle_1$  &  $|\{ss\}^{\bar 3}_1 (\bar n \bar c)^3_1\rangle_1$     &   $|\{ss\}^6_0 (\bar n \bar c)^{\bar 6}_1\rangle_1$    \\
& $\frac{1}{2}(2^+)$       &  $|\{ss\}^{\bar 3}_1 (\bar n \bar c)^3_1\rangle_2$     &  $\cdots$   & $\cdots$ \\

$\{ss\}(\bar n \bar b)$ & $\frac{1}{2}(0^+)$       & $|\{ss\}^{\bar 3}_1 (\bar n \bar b)^3_1\rangle_0$     &  $|\{ss\}^6_0 (\bar n \bar b)^{\bar 6}_0\rangle_0$   & $\cdots$     \\
& $\frac{1}{2}(1^+)$      &  $|\{ss\}^{\bar 3}_1 (\bar n \bar b)^3_0\rangle_1$  &  $|\{ss\}^{\bar 3}_1 (\bar n \bar b)^3_1\rangle_1$     &   $|\{ss\}^6_0 (\bar n \bar b)^{\bar 6}_1\rangle_1$    \\
& $\frac{1}{2}(2^+)$       &  $|\{ss\}^{\bar 3}_1 (\bar n \bar b)^3_1\rangle_2$     &  $\cdots$   & $\cdots$ \\\hline

$\{ss\}(\bar s \bar c)$ & $0(0^+)$       & $|\{ss\}^{\bar 3}_1 (\bar s \bar c)^3_1\rangle_0$     &  $|\{ss\}^6_0 (\bar s \bar c)^{\bar 6}_0\rangle_0$   & $\cdots$     \\
& $0(1^+)$      &  $|\{ss\}^{\bar 3}_1 (\bar s \bar c)^3_0\rangle_1$  &  $|\{ss\}^{\bar 3}_1 (\bar s \bar c)^3_1\rangle_1$     &   $|\{ss\}^6_0 (\bar s \bar c)^{\bar 6}_1\rangle_1$    \\
& $0(2^+)$       &  $|\{ss\}^{\bar 3}_1 (\bar s \bar c)^3_1\rangle_2$     &  $\cdots$   & $\cdots$ \\

$\{ss\}(\bar s \bar b)$ & $0(0^+)$       & $|\{ss\}^{\bar 3}_1 (\bar s \bar b)^3_1\rangle_0$     &  $|\{ss\}^6_0 (\bar s \bar b)^{\bar 6}_0\rangle_0$   & $\cdots$     \\
& $0(1^+)$      &  $|\{ss\}^{\bar 3}_1 (\bar s \bar b)^3_0\rangle_1$  &  $|\{ss\}^{\bar 3}_1 (\bar s \bar b)^3_1\rangle_1$     &   $|\{ss\}^6_0 (\bar s \bar b)^{\bar 6}_1\rangle_1$    \\
& $0(2^+)$       &  $|\{ss\}^{\bar 3}_1 (\bar s \bar b)^3_1\rangle_2$     &  $\cdots$   & $\cdots$ \\\hline\hline

\end{tabular*}
\end{center}
\end{table}

\section{RESULTS AND DISCUSSIONS}{\label{results}}

In this work, we adopt $N^3_{max} = 10^3$ Gaussian bases to calculate the mass spectra of $S-$wave $q_1q_2\bar q_3 \bar Q_4$ tetraquark states systematacially.  Under these large bases,
our numerical results are stable enough for quark model estimations. According to the number of strange quarks, one can classify these open charm and bottom tetraquarks into four groups. We will examine the mass spectra of these systems successively.

The predicted masses for the nonstrange tetraquarks $nn\bar n \bar c$ and $nn\bar n \bar b$ are presented in Table~\ref{mass1}. For the $nn\bar n \bar c$ system, the masses lie in the range of $2570\sim 3327 \rm{MeV}$, while the masses of 
$nn\bar n \bar b$ states vary from 5977 to 6621 MeV. It can be noticed that these mass regions have significant overlap with the excited charmed and bottom mesons. From the Review of Particle Physics~\cite{Tanabashi:2018oca}, there exist several higher charmed and bottom states, which may correspond to the $nn\bar n \bar c$ and $nn\bar n \bar b$ tetraquark states. However, these observed resonances can be described under the conventional interpretations well. In fact, the physical resonances may be the admixtures of the conventional mesons and tetraquarks, which disturbs our understanding. The more efficient way is to hunt for the flavor exotic states $uu \bar d \bar c$, $dd \bar u \bar c$, $uu \bar d \bar b$, and $dd \bar u \bar b$. With large phase space, these flavor exotic states and their antiparticles can easily decay into the conventional charmed or bottom mesons by emitting one or more pions, which can be searched in future experiments.    

\begin{table*}[htp]
\begin{center}
\caption{\label{mass1} Predicted mass spectra for the $nn\bar n \bar c$ and $nn\bar n \bar b$ systems.}
\begin{tabular*}{18cm}{@{\extracolsep{\fill}}p{1.7cm}<{\centering}p{1.7cm}<{\centering}p{6.2cm}<{\centering}p{1.8cm}<{\centering}p{6.2cm}<{\centering}}
\hline\hline
 $I(J^P)$  & Configuration                                             & $\langle H\rangle$ (MeV) & Mass (MeV)  & Eigenvector\\\hline

 $\frac{1}{2}(0^+)$  &  $|[nn]^{\bar 3}_0 (\bar n \bar c)^3_0 \rangle_0$      & \multirow{2}{*}{$\begin{pmatrix}2868&-242 \\-242&2766\end{pmatrix}$}
               & \multirow{2}{*}{$\begin{bmatrix}2570 \\3064 \end{bmatrix}$}  & \multirow{2}{*}{$\begin{bmatrix}(0.630, 0.776)\\(0.776, -0.630)\end{bmatrix}$}\\
               &  $|[nn]^6_1 (\bar n \bar c)^{\bar 6}_1\rangle_0$ \\
 $\frac{1}{2}(1^+)$ &  $|[nn]^{\bar 3}_0 (\bar n \bar c)^3_1\rangle_1$      & \multirow{3}{*}{$\begin{pmatrix}2930&-112&77 \\-112&3129&87 \\ 77& 87 & 2952\end{pmatrix}$}
               & \multirow{3}{*}{$\begin{bmatrix}2802 \\3019 \\ 3190 \end{bmatrix}$}  & \multirow{3}{*}{$\begin{bmatrix}(0.703, 0.397, -0.590)\\(0.632, 0.032, 0.774) \\ (-0.326, 0.917, 0.229)\end{bmatrix}$}\\
                &  $|[nn]^6_1 (\bar n \bar c)^{\bar 6}_0\rangle_1$     \\
              &  $|[nn]^6_1 (\bar n \bar c)^{\bar 6}_1\rangle_1$ \\
$\frac{1}{2}(2^+)$ &   $|[nn]^6_1 (\bar n \bar c)^{\bar 6}_1\rangle_2$    & 3240  &  3240  &  1\\

 $\frac{1}{2}/\frac{3}{2}(0^+)$  & $|\{nn\}^{\bar 3}_1 (\bar n \bar c)^3_1\rangle_0$    & \multirow{2}{*}{$\begin{pmatrix}3050&193 \\193&3192\end{pmatrix}$}
               & \multirow{2}{*}{$\begin{bmatrix}2915 \\3327 \end{bmatrix}$}  & \multirow{2}{*}{$\begin{bmatrix}(-0.820, 0.572)\\(-0.572, -0.820)\end{bmatrix}$}\\
                 &  $|\{nn\}^6_0 (\bar n \bar c)^{\bar 6}_0\rangle_0$    \\
 $\frac{1}{2}/\frac{3}{2}(1^+)$  &  $|\{nn\}^{\bar 3}_1 (\bar n \bar c)^3_0\rangle_1$     & \multirow{3}{*}{$\begin{pmatrix}3100&-38&108 \\-38&3107&70 \\ 108& 70 & 3173\end{pmatrix}$}
               & \multirow{3}{*}{$\begin{bmatrix}2980 \\3140 \\ 3260 \end{bmatrix}$}  & \multirow{3}{*}{$\begin{bmatrix}(0.662, 0.504, -0.555)\\(0.560, -0.824, -0.081) \\ (0.498, 0.257, 0.828)\end{bmatrix}$}\\
                &  $|\{nn\}^{\bar 3}_1 (\bar n \bar c)^3_1\rangle_1$     \\
              &   $|\{nn\}^6_0 (\bar n \bar c)^{\bar 6}_1\rangle_1$     \\
$\frac{1}{2}/\frac{3}{2}(2^+)$    &  $|\{nn\}^{\bar 3}_1 (\bar n \bar c)^3_1\rangle_2$    & 3210  &  3210  &  1\\\hline

 $\frac{1}{2}(0^+)$  &  $|[nn]^{\bar 3}_0 (\bar n \bar b)^3_0 \rangle_0$      & \multirow{2}{*}{$\begin{pmatrix}6206&-201 \\-201&6153\end{pmatrix}$}
               & \multirow{2}{*}{$\begin{bmatrix}5977 \\6382 \end{bmatrix}$}  & \multirow{2}{*}{$\begin{bmatrix}(0.659, 0.752)\\(0.752, -0.659)\end{bmatrix}$}\\
               &  $|[nn]^6_1 (\bar n \bar b)^{\bar 6}_1\rangle_0$ \\
 $\frac{1}{2}(1^+)$ &  $|[nn]^{\bar 3}_0 (\bar n \bar b)^3_1\rangle_1$      & \multirow{3}{*}{$\begin{pmatrix}6230&-97&-113 \\-97&6448&-136 \\ -113& -136 & 6308\end{pmatrix}$}
               & \multirow{3}{*}{$\begin{bmatrix}6080 \\6373 \\ 6534 \end{bmatrix}$}  & \multirow{3}{*}{$\begin{bmatrix}(-0.701, -0.402, -0.589)\\(-0.706, 0.275, 0.653) \\ (0.101, -0.874, 0.476)\end{bmatrix}$}\\
                &  $|[nn]^6_1 (\bar n \bar b)^{\bar 6}_0\rangle_1$     \\
              &  $|[nn]^6_1 (\bar n \bar b)^{\bar 6}_1\rangle_1$ \\
$\frac{1}{2}(2^+)$ &   $|[nn]^6_1 (\bar n \bar b)^{\bar 6}_1\rangle_2$    & 6552  &  6552  &  1\\

 $\frac{1}{2}/\frac{3}{2}(0^+)$  & $|\{nn\}^{\bar 3}_1 (\bar n \bar b)^3_1\rangle_0$    & \multirow{2}{*}{$\begin{pmatrix}6366&167 \\167&6512\end{pmatrix}$}
               & \multirow{2}{*}{$\begin{bmatrix}6256 \\6621 \end{bmatrix}$}  & \multirow{2}{*}{$\begin{bmatrix}(-0.836, 0.548)\\(-0.548, -0.836)\end{bmatrix}$}\\
                 &  $|\{nn\}^6_0 (\bar n \bar b)^{\bar 6}_0\rangle_0$    \\
 $\frac{1}{2}/\frac{3}{2}(1^+)$  &  $|\{nn\}^{\bar 3}_1 (\bar n \bar b)^3_0\rangle_1$     & \multirow{3}{*}{$\begin{pmatrix}6437&-51&92 \\-51&6415&102 \\ 92& 102 & 6503\end{pmatrix}$}
               & \multirow{3}{*}{$\begin{bmatrix}6286 \\6478 \\ 6591 \end{bmatrix}$}  & \multirow{3}{*}{$\begin{bmatrix}(0.546, 0.643, -0.536)\\(0.745, -0.666, -0.039) \\ (0.382, 0.378, 0.843)\end{bmatrix}$}\\
                &  $|\{nn\}^{\bar 3}_1 (\bar n \bar b)^3_1\rangle_1$     \\
              &   $|\{nn\}^6_0 (\bar n \bar b)^{\bar 6}_1\rangle_1$     \\
$\frac{1}{2}/\frac{3}{2}(2^+)$    &  $|\{nn\}^{\bar 3}_1 (\bar n \bar b)^3_1\rangle_2$    & 6503  &  6503  &  1\\

\hline\hline
\end{tabular*}
\end{center}
\end{table*} 

There exist several types of flavor contents for the tetraquark states including one strange quark, and the calculated mass spectra are shown in Table.~\ref{mass2}. Given the $D^-K^+$ decay mode, the newly observed $X_0(2900)$ and $X_1(2900)$ should belong to the $ud \bar s \bar c$ states, and their isospins can be either 0 or 1. From Table.~\ref{mass2}, it can be seen that the predicted masses of $0(0^+)$ states are 2765 and 3125 MeV, where the large splitting arises from the significant mixing scheme of pure $|\bar 3 3\rangle$ and $|6 \bar 6\rangle$ states. Also, the mass of the lowest $1(0^+)$ $nn \bar s \bar c$ state is 3065 MeV, which is larger than the experimental data. Our results disfavor the observed $X_0(2900)$ as a compact $ud \bar s \bar c$ tetraquark. Since the parity of $X_1(2900)$ is negative, it has one orbital excitation at least. From our calculations of $S-$wave states, the $P-$wave $nn \bar s \bar c$ states should have rather large masses, which excludes the assignment of $X_1(2900)$ as a $nn \bar s \bar c$ compact tetraquark state. Other interpretations, such as molecules and kinematic effects, are possible for these two states. In Ref.~\cite{Cheng:2020nho}, the authors adopt the color-magnetic interaction model to obtain four $0^+$ $cs\bar u \bar d$ states with masses of 2320, 2607, 2850, and 3129 MeV, respectively. It seems that the $X_0(2900)$ may be assigned as a higher $0^+$ compact tetraquark through it mass, but the predicted decay width is significant smaller than experimental data. In Ref.~\cite{Karliner:2020vsi}, the lowest $0^+$ $ud \bar s \bar c$ state is estimated to be $2754~\rm{MeV}$ in baryonic-quark picture or $2863~\rm{MeV}$ in the string-junction picture within the simple quark model, where the $X_0(2900)$ can be regarded as a compact tetraquark. The color-magnetic interaction model and simple quark model also disfavor the $X_1(2900)$ as a compact tetraquark state, which is consistent with our calculations. The differences among these works arise from the different choices of interactions. It should be mentioned that a unified treatment of mesons and tetraquarks are essential to obtain the reliable mass spectra of open charm and bottom tetraquarks, and further investigations with various approaches are encouraged. 

\begin{table*}[htp]
\begin{center}
\caption{\label{mass2} Predicted mass spectra for the $nn\bar s \bar c$, $nn\bar s \bar b$, $ns\bar n \bar c$, and $ns\bar n \bar b$  systems.}
\begin{tabular*}{18cm}{@{\extracolsep{\fill}}p{1.7cm}<{\centering}p{1.7cm}<{\centering}p{6.2cm}<{\centering}p{1.8cm}<{\centering}p{6.2cm}<{\centering}}
\hline\hline
 $I(J^P)$  & Configuration                                             & $\langle H\rangle$ (MeV) & Mass (MeV)  & Eigenvector\\\hline

 $0(0^+)$  &  $|[nn]^{\bar 3}_0 (\bar s \bar c)^3_0 \rangle_0$      & \multirow{2}{*}{$\begin{pmatrix}2975&193 \\193&2942\end{pmatrix}$}
               & \multirow{2}{*}{$\begin{bmatrix}2765 \\3152 \end{bmatrix}$}  & \multirow{2}{*}{$\begin{bmatrix}(-0.677, 0.736)\\(0.736, 0.677)\end{bmatrix}$}\\
               &  $|[nn]^6_1 (\bar s \bar c)^{\bar 6}_1\rangle_0$ \\
 $0(1^+)$ &  $|[nn]^{\bar 3}_0 (\bar s \bar c)^3_1\rangle_1$      & \multirow{3}{*}{$\begin{pmatrix}3027&91&43 \\91&3223&-44 \\ 43& -44 & 3085\end{pmatrix}$}
               & \multirow{3}{*}{$\begin{bmatrix}2964 \\3108 \\ 3263 \end{bmatrix}$}  & \multirow{3}{*}{$\begin{bmatrix}(-0.827, 0.364, 0.428)\\(0.455, -0.015, 0.891) \\ (-0.330, -0.931, 0.153)\end{bmatrix}$}\\
                &  $|[nn]^6_1 (\bar s \bar c)^{\bar 6}_0\rangle_1$     \\
              &  $|[nn]^6_1 (\bar s \bar c)^{\bar 6}_1\rangle_1$ \\
$0(2^+)$ &   $|[nn]^6_1 (\bar s \bar c)^{\bar 6}_1\rangle_2$    & 3316  &  3316  &  1\\

 $1(0^+)$  & $|\{nn\}^{\bar 3}_1 (\bar s \bar c)^3_1\rangle_0$    & \multirow{2}{*}{$\begin{pmatrix}3172&-154 \\-154&3289\end{pmatrix}$}
               & \multirow{2}{*}{$\begin{bmatrix}3065 \\3396 \end{bmatrix}$}  & \multirow{2}{*}{$\begin{bmatrix}(-0.823, -0.568)\\(0.568, -0.823)\end{bmatrix}$}\\
                 &  $|\{nn\}^6_0 (\bar s \bar c)^{\bar 6}_0\rangle_0$    \\
 $1(1^+)$  &  $|\{nn\}^{\bar 3}_1 (\bar s \bar c)^3_0\rangle_1$     & \multirow{3}{*}{$\begin{pmatrix}3211&23&87 \\23&3218&-40 \\ 87& -40 & 3275\end{pmatrix}$}
               & \multirow{3}{*}{$\begin{bmatrix}3130 \\3235 \\ 3339 \end{bmatrix}$}  & \multirow{3}{*}{$\begin{bmatrix}(-0.713, 0.438, 0.548)\\(-0.457, -0.883, 0.110) \\ (-0.532, 0.171, -0.829)\end{bmatrix}$}\\
                &  $|\{nn\}^{\bar 3}_1 (\bar s \bar c)^3_1\rangle_1$     \\
              &   $|\{nn\}^6_0 (\bar s \bar c)^{\bar 6}_1\rangle_1$     \\
$1(2^+)$    &  $|\{nn\}^{\bar 3}_1 (\bar s \bar c)^3_1\rangle_2$    & 3302  &  3302  &  1\\\hline

 $0(0^+)$  &  $|[nn]^{\bar 3}_0 (\bar s \bar b)^3_0 \rangle_0$      & \multirow{2}{*}{$\begin{pmatrix}6315&147 \\147&6372\end{pmatrix}$}
               & \multirow{2}{*}{$\begin{bmatrix}6194 \\6493 \end{bmatrix}$}  & \multirow{2}{*}{$\begin{bmatrix}(-0.771, 0.636)\\(-0.636, -0.771)\end{bmatrix}$}\\
               &  $|[nn]^6_1 (\bar s \bar b)^{\bar 6}_1\rangle_0$ \\
 $0(1^+)$ &  $|[nn]^{\bar 3}_0 (\bar s \bar b)^3_1\rangle_1$      & \multirow{3}{*}{$\begin{pmatrix}6336&-72&-73 \\-72&6580&-84 \\ -73& -84 & 6479\end{pmatrix}$}
               & \multirow{3}{*}{$\begin{bmatrix}6272 \\6492 \\ 6630 \end{bmatrix}$}  & \multirow{3}{*}{$\begin{bmatrix}(-0.848, -0.314, -0.427)\\(-0.519, 0.331, 0.788) \\ (0.107, -0.890, 0.443)\end{bmatrix}$}\\
                &  $|[nn]^6_1 (\bar s \bar b)^{\bar 6}_0\rangle_1$     \\
              &  $|[nn]^6_1 (\bar s \bar b)^{\bar 6}_1\rangle_1$ \\
$0(2^+)$ &   $|[nn]^6_1 (\bar s \bar b)^{\bar 6}_1\rangle_2$    & 6656  &  6656  &  1\\

 $1(0^+)$  & $|\{nn\}^{\bar 3}_1 (\bar s \bar b)^3_1\rangle_0$    & \multirow{2}{*}{$\begin{pmatrix}6499&122 \\122&6645\end{pmatrix}$}
               & \multirow{2}{*}{$\begin{bmatrix}6430 \\6714 \end{bmatrix}$}  & \multirow{2}{*}{$\begin{bmatrix}(-0.870, 0.493)\\(-0.493, -0.870)\end{bmatrix}$}\\
                 &  $|\{nn\}^6_0 (\bar s \bar b)^{\bar 6}_0\rangle_0$    \\
 $1(1^+)$  &  $|\{nn\}^{\bar 3}_1 (\bar s \bar b)^3_0\rangle_1$     & \multirow{3}{*}{$\begin{pmatrix}6549&35&68 \\35&6535&-66 \\ 68& -66 & 6639\end{pmatrix}$}
               & \multirow{3}{*}{$\begin{bmatrix}6457 \\6577 \\ 6689 \end{bmatrix}$}  & \multirow{3}{*}{$\begin{bmatrix}(0.592, -0.659, -0.463)\\(-0.722, -0.689, 0.059) \\ (0.358, -0.300, 0.884)\end{bmatrix}$}\\
                &  $|\{nn\}^{\bar 3}_1 (\bar s \bar b)^3_1\rangle_1$     \\
              &   $|\{nn\}^6_0 (\bar s \bar b)^{\bar 6}_1\rangle_1$     \\
$1(2^+)$    &  $|\{nn\}^{\bar 3}_1 (\bar s \bar b)^3_1\rangle_2$    & 6602  &  6602  &  1\\\hline

 $0/1(0^+)$  &  $|[ns]^{\bar 3}_0 (\bar n \bar c)^3_0 \rangle_0$      & \multirow{2}{*}{$\begin{pmatrix}3071&-201 \\-201&2960\end{pmatrix}$}
               & \multirow{2}{*}{$\begin{bmatrix}2807 \\3224 \end{bmatrix}$}  & \multirow{2}{*}{$\begin{bmatrix}(0.606, 0.795)\\(0.795, -0.606)\end{bmatrix}$}\\
               &  $|[ns]^6_1 (\bar n \bar c)^{\bar 6}_1\rangle_0$ \\
 $0/1(1^+)$ &  $|[ns]^{\bar 3}_0 (\bar n \bar c)^3_1\rangle_1$      & \multirow{3}{*}{$\begin{pmatrix}3133&95&60 \\95&3270&-68 \\ 60& -68 & 3116\end{pmatrix}$}
               & \multirow{3}{*}{$\begin{bmatrix}3013 \\3181 \\ 3325 \end{bmatrix}$}  & \multirow{3}{*}{$\begin{bmatrix}(0.645, -0.409, -0.646)\\(-0.656, 0.138, -0.742) \\ (-0.392, -0.902, 0.180)\end{bmatrix}$}\\
                &  $|[ns]^6_1 (\bar n \bar c)^{\bar 6}_0\rangle_1$     \\
              &  $|[ns]^6_1 (\bar n \bar c)^{\bar 6}_1\rangle_1$ \\
$0/1(2^+)$ &   $|[ns]^6_1 (\bar n \bar c)^{\bar 6}_1\rangle_2$    & 3364  &  3364  &  1\\

 $0/1(0^+)$  & $|\{ns\}^{\bar 3}_1 (\bar n \bar c)^3_1\rangle_0$    & \multirow{2}{*}{$\begin{pmatrix}3205&-165 \\-165&3317\end{pmatrix}$}
               & \multirow{2}{*}{$\begin{bmatrix}3087 \\3435 \end{bmatrix}$}  & \multirow{2}{*}{$\begin{bmatrix}(-0.813, -0.583)\\(0.583, -0.813)\end{bmatrix}$}\\
                 &  $|\{ns\}^6_0 (\bar n \bar c)^{\bar 6}_0\rangle_0$    \\
 $0/1(1^+)$  &  $|\{ns\}^{\bar 3}_1 (\bar n \bar c)^3_0\rangle_1$     & \multirow{3}{*}{$\begin{pmatrix}3236&30&93 \\30&3253&-56 \\ 93& -56 & 3297\end{pmatrix}$}
               & \multirow{3}{*}{$\begin{bmatrix}3139 \\3275 \\ 3372 \end{bmatrix}$}  & \multirow{3}{*}{$\begin{bmatrix}(0.685, -0.459, -0.566)\\(-0.523, -0.851, 0.057) \\ (0.507, -0.256, 0.823)\end{bmatrix}$}\\
                &  $|\{ns\}^{\bar 3}_1 (\bar n \bar c)^3_1\rangle_1$     \\
              &   $|\{ns\}^6_0 (\bar n \bar c)^{\bar 6}_1\rangle_1$     \\
$0/1(2^+)$    &  $|\{ns\}^{\bar 3}_1 (\bar n \bar c)^3_1\rangle_2$    & 3339  &  3339  &  1\\\hline

 $0/1(0^+)$  &  $|[ns]^{\bar 3}_0 (\bar n \bar b)^3_0 \rangle_0$      & \multirow{2}{*}{$\begin{pmatrix}6406&164 \\164&6336\end{pmatrix}$}
               & \multirow{2}{*}{$\begin{bmatrix}6203 \\6538 \end{bmatrix}$}  & \multirow{2}{*}{$\begin{bmatrix}(-0.629, 0.778)\\(0.778, 0.629)\end{bmatrix}$}\\
               &  $|[ns]^6_1 (\bar n \bar b)^{\bar 6}_1\rangle_0$ \\
 $0/1(1^+)$ &  $|[ns]^{\bar 3}_0 (\bar n \bar b)^3_1\rangle_1$      & \multirow{3}{*}{$\begin{pmatrix}6430&81&91 \\81&6581&-110 \\ 91& -110 & 6462\end{pmatrix}$}
               & \multirow{3}{*}{$\begin{bmatrix}6292 \\6531 \\ 6650 \end{bmatrix}$}  & \multirow{3}{*}{$\begin{bmatrix}(-0.657, 0.422, 0.624)\\(0.740, 0.209, 0.639) \\ (-0.139, -0.882, 0.450)\end{bmatrix}$}\\
                &  $|[ns]^6_1 (\bar n \bar b)^{\bar 6}_0\rangle_1$     \\
              &  $|[ns]^6_1 (\bar n \bar b)^{\bar 6}_1\rangle_1$ \\
$0/1(2^+)$ &   $|[ns]^6_1 (\bar n \bar b)^{\bar 6}_1\rangle_2$    & 6668  &  6668  &  1\\

 $0/1(0^+)$  & $|\{ns\}^{\bar 3}_1 (\bar n \bar b)^3_1\rangle_0$    & \multirow{2}{*}{$\begin{pmatrix}6516&140 \\140&6628\end{pmatrix}$}
               & \multirow{2}{*}{$\begin{bmatrix}6421 \\6723 \end{bmatrix}$}  & \multirow{2}{*}{$\begin{bmatrix}(-0.828, 0.560)\\(-0.560, -0.828)\end{bmatrix}$}\\
                 &  $|\{ns\}^6_0 (\bar n \bar b)^{\bar 6}_0\rangle_0$    \\
 $0/1(1^+)$  &  $|\{ns\}^{\bar 3}_1 (\bar n \bar b)^3_0\rangle_1$     & \multirow{3}{*}{$\begin{pmatrix}6570&-41&78 \\-41&6556&83 \\78& 83 & 6619\end{pmatrix}$}
               & \multirow{3}{*}{$\begin{bmatrix}6446 \\6605 \\ 6694 \end{bmatrix}$}  & \multirow{3}{*}{$\begin{bmatrix}(0.554, 0.625, -0.549)\\(0.732, -0.680, -0.037) \\ (0.397, 0.382, 0.835)\end{bmatrix}$}\\
                &  $|\{ns\}^{\bar 3}_1 (\bar n \bar b)^3_1\rangle_1$     \\
              &   $|\{ns\}^6_0 (\bar n \bar b)^{\bar 6}_1\rangle_1$     \\
$0/1(2^+)$    &  $|\{ns\}^{\bar 3}_1 (\bar n \bar b)^3_1\rangle_2$    & 6630  &  6630  &  1\\

\hline\hline
\end{tabular*}
\end{center}
\end{table*}

\begin{table*}[htp]
\begin{center}
\caption{\label{pro} The color proportions and root mean square radii of the lower $nn\bar s \bar c$ states. The expectations $\langle \boldsymbol r_{14}^2  \rangle^{1/2}$, $\langle \boldsymbol r_{23}^2
\rangle^{1/2}$, and $\langle \boldsymbol r^{\prime \prime 2} \rangle^{1/2}$ equal to the values of $\langle
\boldsymbol r_{24}^2  \rangle^{1/2}$, $\langle \boldsymbol r_{13}^2  \rangle^{1/2}$, and $\langle \boldsymbol r^{\prime 2} \rangle^{1/2}$,
respectively, which are omitted for simplicity. The units of masses and root mean square radii are in MeV and fm, respectively.}
\begin{tabular*}{18cm}{@{\extracolsep{\fill}}*{10}{p{1.3cm}<{\centering}}}
\hline\hline
 $I(J^P)$  & Mass  &   $|\bar 3 3\rangle$  &  $|6 \bar 6\rangle$  &    $\langle \boldsymbol r_{12}^2
 \rangle^{1/2}$    &  $\langle \boldsymbol r_{34}^2 \rangle^{1/2}$   &  $\langle \boldsymbol r^2 \rangle^{1/2}$  &
 $\langle \boldsymbol r_{13}^2  \rangle^{1/2}$   &  $\langle \boldsymbol r_{24}^2  \rangle^{1/2}$
 &  $\langle \boldsymbol r^{\prime 2} \rangle^{1/2}$  \\\hline
 $0(0^+)$  & 2765  &   45.8\%  &  54.2\%    &  0.524  &  0.480     &  0.351  &  0.581  &  0.449
 &  0.367 \\
 $0(1^+)$  & 2964  &   68.4\%  &  31.6\%   &  0.533  &  0.520     &  0.388  &  0.627  &  0.482
 &  0.395 \\
 $0(2^+)$  & 3316  &   0\%  &  100\%   &  0.688  &  0.640     &  0.344  &  0.704  &  0.504
 &  0.483 \\
 $1(0^+)$  & 3065  &   67.7\%  &  32.3\%   &  0.616  &  0.529     &  0.389  &  0.650  &  0.507
 &  0.407 \\
 $1(1^+)$  & 3130  &   70.0\%  &  30.0\%    &  0.622  &  0.521     &  0.407  &  0.659  &  0.523
 &  0.404 \\
 $1(2^+)$  & 3302  &   100\%  &  0\%    &  0.603  &  0.518     &  0.457  &  0.685  &  0.558
 &  0.403 \\
\hline\hline
\end{tabular*}
\end{center}
\end{table*} 

\begin{figure}[!htb]
\includegraphics[scale=1.0]{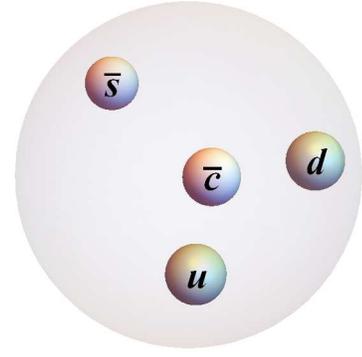}
\vspace{0.0cm} \caption{The sketch of lower $ud \bar s \bar c$ states.}
\label{udsc}
\end{figure}

The predicted color proportions and root mean square radii of the lower $nn \bar s \bar c$ states are listed in Table~\ref{pro}. Our results show that these states have relatively small root mean square radii, which indicates that all of them have compact inner structures. The sketch of these $ud \bar s \bar c$ tetraquarks is also plotted in Fig.~\ref{udsc}. It can be seen that the four quarks are separated from each other in a compact tetraquark, which is quite different with the diquark-antidiquark or loosely bound molecular picture.

The $nn \bar s \bar b$ states are the bottom partners of $nn \bar s \bar c$ states, and all the predicted masses lie above the respective thresholds. In Ref.~\cite{Yu:2017pmn}, the author proposed a possible stable  $bs \bar u \bar d$ state, while the calculations under potential model indicate that no stable diquark-antidiquark $bs \bar u \bar d$ state exists~\cite{Chen:2018hts}. Also, the color-magnetic interaction model suggest that the lowest $0^+$ and $1^+$ compact $bs \bar u \bar d$ states should be near the relevant thresholds~\cite{Cheng:2020nho}. More theoretical and experimental efforts are needed to clarify this problem.

Unlike the $nn \bar s \bar c$ and $nn \bar s \bar b$ states, some of the $ns \bar n \bar c$ and $ns \bar n \bar b$ states can mix with the conventional charmed-strange and bottom-strange mesons. At the early stage, the investigations on $ns \bar n \bar c$ and $ns \bar n \bar b$ states mainly focus on the possible tetraquark interpretation of the $D^*_{s0}(2317)$. Since the light quark pair can create or annihilate easily, it is difficult to distinguish various explanations of $D^*_{s0}(2317)$. Hence, searching for the flavor exotic states $us \bar d \bar c$, $ds \bar u \bar c$, $us \bar d \bar b$, and $ds \bar u \bar b$ seems to be more worthwhile. After the observation of $X(5568)$, A plenty of studies on the $ns \bar n \bar b$ system have been performed. Our results on $ns \bar n \bar c$ and $ns \bar n \bar b$ states are much higher than the masses of $D^*_{s0}(2317)$ and $X(5568)$, which exclude their tetraquark interpretations.

For the tetraquarks $nn \bar s \bar c$ and $nn \bar s \bar b$, they can decay into the $\bar D^{(*)} K^{(*)}$ and $B^{(*)} K^{(*)}$ final states via fall apart mechanism, respectively. For the $ns \bar n \bar c$ and $ns \bar n \bar b$ states, the possible decay channels are $\bar D^{(*)} \bar K^{(*)}$, $D_s^{(*)-} \pi$ and $B^{(*)} \bar K^{(*)}$ and $B_s^{(*)} \pi$. Certainly, the antiparticles of these tetraquarks decay into the similar final states under charge conjugate transformation. We hope our results can provide helpful information for hunting for these flavor exotic tetraquarks in future LHCb and BelleII experiments.

The masses of tetraquark states including two strange quarks are listed in Table.~\ref{mass3}. For the $ns \bar s \bar c$ and  $ns \bar s \bar b$ states, the predicted masses are much higher than the relevant thresholds, which may fall apart easily. Also, these states can mix with the conventional charmed and bottom mesons, and be hardly picked out from the excited mesons. The more interesting systems are $ss \bar n \bar c$ and  $ss \bar n \bar b$ states, which are totally flavor exotic. They and their antiparticles can be searched in the $D_s^{(*)-} \bar K^{(*)}$, $D_s^{(*)+} K^{(*)}$, $B_s^{(*)} \bar K^{(*)}$, and $\bar B_s^{(*)} K^{(*)}$ final states by future experiments.

\begin{table*}[!htp]
\begin{center}
\caption{\label{mass3} Predicted mass spectra for the $ns\bar s \bar c$, $ns\bar s \bar b$, $ss\bar n \bar c$, and $ss\bar n \bar b$  systems.}
\begin{tabular*}{18cm}{@{\extracolsep{\fill}}p{1.7cm}<{\centering}p{1.7cm}<{\centering}p{6.2cm}<{\centering}p{1.8cm}<{\centering}p{6.2cm}<{\centering}}
\hline\hline
 $I(J^P)$  & Configuration                                             & $\langle H\rangle$ (MeV) & Mass (MeV)  & Eigenvector\\\hline

 $\frac{1}{2}(0^+)$  &  $|[ns]^{\bar 3}_0 (\bar s \bar c)^3_0 \rangle_0$      & \multirow{2}{*}{$\begin{pmatrix}3175&168 \\168&3104\end{pmatrix}$}
               & \multirow{2}{*}{$\begin{bmatrix}2967 \\3312 \end{bmatrix}$}  & \multirow{2}{*}{$\begin{bmatrix}(-0.630, 0.776)\\(0.776, 0.630)\end{bmatrix}$}\\
               &  $|[ns]^6_1 (\bar s \bar c)^{\bar 6}_1\rangle_0$ \\
 $\frac{1}{2}(1^+)$ &  $|[ns]^{\bar 3}_0 (\bar s \bar c)^3_1\rangle_1$      & \multirow{3}{*}{$\begin{pmatrix}3227&-80&36 \\-80&3357&37 \\ 36& 37 & 3231\end{pmatrix}$}
               & \multirow{3}{*}{$\begin{bmatrix}3156 \\3261 \\ 3397 \end{bmatrix}$}  & \multirow{3}{*}{$\begin{bmatrix}(0.732, 0.397, -0.553)\\(0.548, 0.139, 0.825) \\ (-0.404, 0.907, 0.116)\end{bmatrix}$}\\
                &  $|[ns]^6_1 (\bar s \bar c)^{\bar 6}_0\rangle_1$     \\
              &  $|[ns]^6_1 (\bar s \bar c)^{\bar 6}_1\rangle_1$ \\
$\frac{1}{2}(2^+)$ &   $|[ns]^6_1 (\bar s \bar c)^{\bar 6}_1\rangle_2$    & 3438  &  3438  &  1\\

 $\frac{1}{2}(0^+)$  & $|\{ns\}^{\bar 3}_1 (\bar s \bar c)^3_1\rangle_0$    & \multirow{2}{*}{$\begin{pmatrix}3317&138 \\138&3406\end{pmatrix}$}
               & \multirow{2}{*}{$\begin{bmatrix}3217 \\3506 \end{bmatrix}$}  & \multirow{2}{*}{$\begin{bmatrix}(-0.808, 0.589)\\(-0.589, -0.808)\end{bmatrix}$}\\
                 &  $|\{ns\}^6_0 (\bar s \bar c)^{\bar 6}_0\rangle_0$    \\
 $\frac{1}{2}(1^+)$  &  $|\{ns\}^{\bar 3}_1 (\bar s \bar c)^3_0\rangle_1$     & \multirow{3}{*}{$\begin{pmatrix}3344&20&78 \\20&3357&-34 \\ 78& -34 & 3391\end{pmatrix}$}
               & \multirow{3}{*}{$\begin{bmatrix}3271 \\3370 \\ 3451 \end{bmatrix}$}  & \multirow{3}{*}{$\begin{bmatrix}(0.720, -0.389, -0.575)\\(-0.417, -0.905, 0.090) \\ (-0.555, 0.175, -0.813)\end{bmatrix}$}\\
                &  $|\{ns\}^{\bar 3}_1 (\bar s \bar c)^3_1\rangle_1$     \\
              &   $|\{ns\}^6_0 (\bar s \bar c)^{\bar 6}_1\rangle_1$     \\
$\frac{1}{2}(2^+)$    &  $|\{ns\}^{\bar 3}_1 (\bar s \bar c)^3_1\rangle_2$    & 3431  &  3431  &  1\\\hline

 $\frac{1}{2}(0^+)$  &  $|[ns]^{\bar 3}_0 (\bar s \bar b)^3_0 \rangle_0$      & \multirow{2}{*}{$\begin{pmatrix}6511&127 \\127&6520\end{pmatrix}$}
               & \multirow{2}{*}{$\begin{bmatrix}6388 \\6643 \end{bmatrix}$}  & \multirow{2}{*}{$\begin{bmatrix}(-0.719, 0.695)\\(-0.695, -0.719)\end{bmatrix}$}\\
               &  $|[ns]^6_1 (\bar s \bar b)^{\bar 6}_1\rangle_0$ \\
 $\frac{1}{2}(1^+)$ &  $|[ns]^{\bar 3}_0 (\bar s \bar b)^3_1\rangle_1$      & \multirow{3}{*}{$\begin{pmatrix}6532&-63&-62 \\-63&6705&-72 \\ -62& -72 & 6614\end{pmatrix}$}
               & \multirow{3}{*}{$\begin{bmatrix}6464 \\6639 \\ 6748 \end{bmatrix}$}  & \multirow{3}{*}{$\begin{bmatrix}(-0.789, -0.357, -0.500)\\(-0.598, 0.261, 0.758) \\ (0.141, -0.897, 0.419)\end{bmatrix}$}\\
                &  $|[ns]^6_1 (\bar s \bar b)^{\bar 6}_0\rangle_1$     \\
              &  $|[ns]^6_1 (\bar s \bar b)^{\bar 6}_1\rangle_1$ \\
$\frac{1}{2}(2^+)$ &   $|[ns]^6_1 (\bar s \bar b)^{\bar 6}_1\rangle_2$    & 6772 &  6772  &  1\\

 $\frac{1}{2}(0^+)$  & $|\{ns\}^{\bar 3}_1 (\bar s \bar b)^3_1\rangle_0$    & \multirow{2}{*}{$\begin{pmatrix}6638&-108 \\-108&6753\end{pmatrix}$}
               & \multirow{2}{*}{$\begin{bmatrix}6573 \\6818 \end{bmatrix}$}  & \multirow{2}{*}{$\begin{bmatrix}(-0.858, -0.513)\\(0.513, -0.858)\end{bmatrix}$}\\
                 &  $|\{ns\}^6_0 (\bar s \bar b)^{\bar 6}_0\rangle_0$    \\
 $\frac{1}{2}(1^+)$  &  $|\{ns\}^{\bar 3}_1 (\bar s \bar b)^3_0\rangle_1$     & \multirow{3}{*}{$\begin{pmatrix}6679&-30&-60 \\-30&6669&-57 \\ -60& -57 & 6747\end{pmatrix}$}
               & \multirow{3}{*}{$\begin{bmatrix}6598 \\6704 \\ 6793 \end{bmatrix}$}  & \multirow{3}{*}{$\begin{bmatrix}(0.597, 0.637, 0.487)\\(0.706, -0.706, 0.057) \\ (-0.379, -0.310, 0.872)\end{bmatrix}$}\\
                &  $|\{ns\}^{\bar 3}_1 (\bar s \bar b)^3_1\rangle_1$     \\
              &   $|\{ns\}^6_0 (\bar s \bar b)^{\bar 6}_1\rangle_1$     \\
$\frac{1}{2}(2^+)$    &  $|\{ns\}^{\bar 3}_1 (\bar s \bar b)^3_1\rangle_2$    & 6728  &  6728  &  1\\\hline

 $\frac{1}{2}(0^+)$  & $|\{ss\}^{\bar 3}_1 (\bar n \bar c)^3_1\rangle_0$    & \multirow{2}{*}{$\begin{pmatrix}3326&-144 \\-144&3413\end{pmatrix}$}
               & \multirow{2}{*}{$\begin{bmatrix}3219 \\3520 \end{bmatrix}$}  & \multirow{2}{*}{$\begin{bmatrix}(-0.802, -0.570)\\(0.570, -0.802)\end{bmatrix}$}\\
                 &  $|\{ss\}^6_0 (\bar n \bar c)^{\bar 6}_0\rangle_0$    \\
 $\frac{1}{2}(1^+)$  &  $|\{ss\}^{\bar 3}_1 (\bar n \bar c)^3_0\rangle_1$     & \multirow{3}{*}{$\begin{pmatrix}3344&-24&81 \\-24&3368&44 \\ 81& 44 & 3392\end{pmatrix}$}
               & \multirow{3}{*}{$\begin{bmatrix}3263 \\3382 \\ 3458 \end{bmatrix}$}  & \multirow{3}{*}{$\begin{bmatrix}(0.705, 0.404, -0.583)\\(-0.475, 0.880, 0.035) \\ (0.527, 0.252, 0.812)\end{bmatrix}$}\\
                &  $|\{ss\}^{\bar 3}_1 (\bar n \bar c)^3_1\rangle_1$     \\
              &   $|\{ss\}^6_0 (\bar n \bar c)^{\bar 6}_1\rangle_1$     \\
$\frac{1}{2}(2^+)$    &  $|\{ss\}^{\bar 3}_1 (\bar n \bar c)^3_1\rangle_2$    & 3443  &  3443  &  1\\\hline

 $\frac{1}{2}(0^+)$  & $|\{ss\}^{\bar 3}_1 (\bar n \bar b)^3_1\rangle_0$    & \multirow{2}{*}{$\begin{pmatrix}6631&-120 \\-120&6716\end{pmatrix}$}
               & \multirow{2}{*}{$\begin{bmatrix}6547 \\6800 \end{bmatrix}$}  & \multirow{2}{*}{$\begin{bmatrix}(-0.816, -0.577)\\(0.577, -0.816)\end{bmatrix}$}\\
                 &  $|\{ss\}^6_0 (\bar n \bar b)^{\bar 6}_0\rangle_0$    \\
 $\frac{1}{2}(1^+)$  &  $|\{ss\}^{\bar 3}_1 (\bar n \bar b)^3_0\rangle_1$     & \multirow{3}{*}{$\begin{pmatrix}6674&-34&67 \\-34&6665&69 \\ 67& 69 & 6707\end{pmatrix}$}
               & \multirow{3}{*}{$\begin{bmatrix}6569 \\6704 \\ 6773 \end{bmatrix}$}  & \multirow{3}{*}{$\begin{bmatrix}(0.559, 0.602, -0.571)\\(0.715, -0.698, -0.035) \\ (0.419, 0.388, 0.821)\end{bmatrix}$}\\
                &  $|\{ss\}^{\bar 3}_1 (\bar n \bar b)^3_1\rangle_1$     \\
              &   $|\{ss\}^6_0 (\bar n \bar b)^{\bar 6}_1\rangle_1$     \\
$\frac{1}{2}(2^+)$    &  $|\{ss\}^{\bar 3}_1 (\bar n \bar b)^3_1\rangle_2$    & 6729  &  6729  &  1\\

\hline\hline
\end{tabular*}
\end{center}
\end{table*}

The calculated masses of $ss \bar s \bar c$ and  $ss \bar s \bar b$  systems are presented in Table.~\ref{mass4}, which lie above 3300 and 6600 MeV, respectively. These states can mix with the conventional charmed-strange and bottom-strange mesons via the strange quark pair annihilation. Current experiments have not investigated these energy regions, and future experimental searches can test our calculations.

\begin{table*}[htp]
\begin{center}
\caption{\label{mass4} Predicted mass spectra for the $ss\bar s \bar c$ and $ss\bar s \bar b$ systems.}
\begin{tabular*}{18cm}{@{\extracolsep{\fill}}p{1.7cm}<{\centering}p{1.7cm}<{\centering}p{6.2cm}<{\centering}p{1.8cm}<{\centering}p{6.2cm}<{\centering}}
\hline\hline
 $I(J^P)$  & Configuration                                             & $\langle H\rangle$ (MeV) & Mass (MeV)  & Eigenvector\\\hline
 $0(0^+)$  & $|\{ss\}^{\bar 3}_1 (\bar s \bar c)^3_1\rangle_0$    & \multirow{2}{*}{$\begin{pmatrix}3429&126 \\126&3494\end{pmatrix}$}
               & \multirow{2}{*}{$\begin{bmatrix}3331 \\3592 \end{bmatrix}$}  & \multirow{2}{*}{$\begin{bmatrix}(-0.790, 0.613)\\(-0.613, -0.790)\end{bmatrix}$}\\
                 &  $|\{ss\}^6_0 (\bar s \bar c)^{\bar 6}_0\rangle_0$    \\
 $0(1^+)$  &  $|\{ss\}^{\bar 3}_1 (\bar s \bar c)^3_0\rangle_1$     & \multirow{3}{*}{$\begin{pmatrix}3448&-17&71 \\-17&3466&29 \\ 71& 29 & 3479\end{pmatrix}$}
               & \multirow{3}{*}{$\begin{bmatrix}3379 \\3475 \\ 3539 \end{bmatrix}$}  & \multirow{3}{*}{$\begin{bmatrix}(0.716, 0.339, -0.610)\\(-0.374, 0.924, 0.075) \\ (0.590, 0.174, 0.789)\end{bmatrix}$}\\
                &  $|\{ss\}^{\bar 3}_1 (\bar s \bar c)^3_1\rangle_1$     \\
              &   $|\{ss\}^6_0 (\bar s \bar c)^{\bar 6}_1\rangle_1$     \\
$0(2^+)$    &  $|\{ss\}^{\bar 3}_1 (\bar s \bar c)^3_1\rangle_2$    & 3533  &  3533  &  1\\\hline

 $0(0^+)$  & $|\{ss\}^{\bar 3}_1 (\bar s \bar b)^3_1\rangle_0$    & \multirow{2}{*}{$\begin{pmatrix}6745&98 \\98&6835\end{pmatrix}$}
               & \multirow{2}{*}{$\begin{bmatrix}6682 \\6898 \end{bmatrix}$}  & \multirow{2}{*}{$\begin{bmatrix}(-0.843, 0.538)\\(-0.538, -0.843)\end{bmatrix}$}\\
                 &  $|\{ss\}^6_0 (\bar s \bar b)^{\bar 6}_0\rangle_0$    \\
 $0(1^+)$  &  $|\{ss\}^{\bar 3}_1 (\bar s \bar b)^3_0\rangle_1$     & \multirow{3}{*}{$\begin{pmatrix}6779&26&-55 \\26&6773&50 \\ -55& 50 & 6828\end{pmatrix}$}
               & \multirow{3}{*}{$\begin{bmatrix}6705 \\6802 \\ 6874 \end{bmatrix}$}  & \multirow{3}{*}{$\begin{bmatrix}(0.598, -0.612, 0.517)\\(-0.690, -0.722, -0.057) \\ (-0.408, 0.322, 0.854)\end{bmatrix}$}\\
                &  $|\{ss\}^{\bar 3}_1 (\bar s \bar b)^3_1\rangle_1$     \\
              &   $|\{ss\}^6_0 (\bar s \bar b)^{\bar 6}_1\rangle_1$     \\
$0(2^+)$    &  $|\{ss\}^{\bar 3}_1 (\bar s \bar b)^3_1\rangle_2$    & 6826  &  6826  &  1\\

\hline\hline
\end{tabular*}
\end{center}
\end{table*}

\begin{figure*}[h]
\includegraphics[scale=0.45]{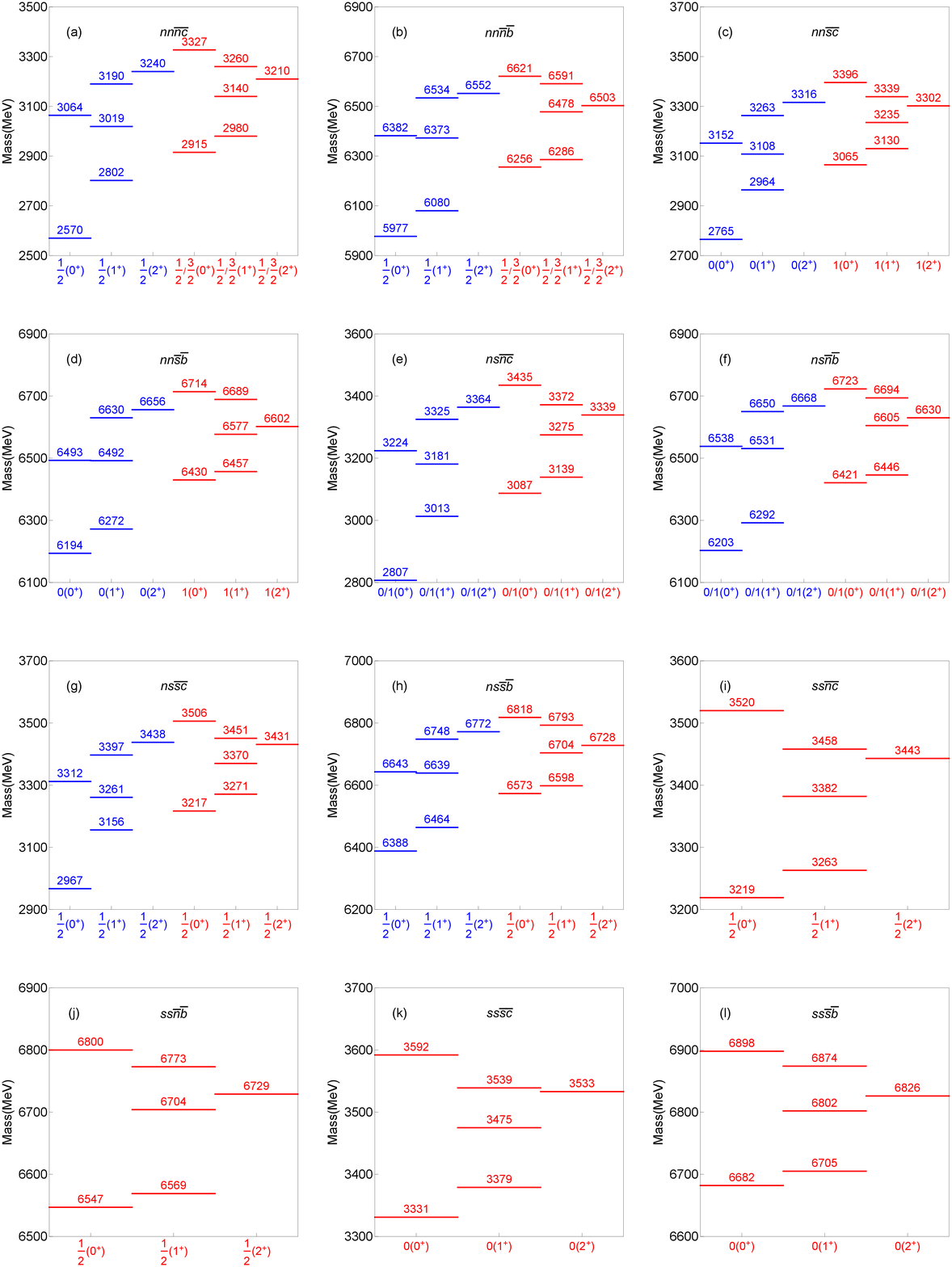}
\vspace{-0.0cm} \caption{The predicted mass spectra of open charm and bottom tetraquarks.}
\label{qqqQ}
\end{figure*}

Finally, we plot the full mass spectra of open charm and bottom tetraquarks in Fig.~\ref{qqqQ}. It can be seen that the spectra for various systems show quite similar patterns, which indicates that the approximate light flavor SU(3) symmetry and heavy quark symmetry are preserved well for the ground states of singly heavy tetraquarks. These two symmetries have achieved great successes for the traditional hadrons, which will also provide a powerful tool for us to investigate the singly heavy tetraquarks. Compared with the prosperities of conventional heavy-light mesons and singly heavy baryons, the studies on singly heavy tetraquarks are far from enough. More theoretical and experimental efforts are encouraged to increase our understanding on these systems.

\section{Summary}{\label{Summary}}

In this work, we systematically investigate the mass spectra of open charm and bottom tetraquarks $q q\bar q \bar Q$ within 
an extended relativized quark model. By using the variational method, the four-body relativized Hamiltonian including the
Coulomb potential, confining potential, spin-spin interactions, and relativistic corrections are solved. The predicted 
masses of four $0^+$ $ud\bar s \bar c$ states are 2765, 3065, 3152, and 3396 MeV, which disfavors the assignment of 
$X_0(2900)$ as a compact tetraquark.

The whole mass spectra of open charm and bottom tetraquark show quite similar patterns, which preserves the light flavor 
SU(3) symmetry and heavy quark symmetry well. Besides the mass spectra, the possible decay modes are also discussed. 
Our results suggest that the future experiments can search for the flavor exotic states $nn\bar s \bar c$, 
$nn\bar s \bar b$, $ss\bar n \bar c$, and $ss\bar n \bar b$ in the heavy-light meson plus kaon final states. More 
theoretical and experimental efforts are needed to investigate these singly heavy tetraquarks.

\bigskip
\noindent
\begin{center}
{\bf ACKNOWLEDGEMENTS}\\
\end{center}
We would like to thank Xian-Hui Zhong for valuable discussions. This project is supported by the National
Natural Science Foundation of China under Grants No.~11705056, No.~11775050, No.~11947224,
No.~11975245, and No.~U1832173, by the fund provided to the Sino-German CRC 110 ``Symmetries and the Emergence of Structure in QCD"
project by the NSFC under Grant No.~11621131001, and by the Key Research Program of Frontier Sciences, CAS, Grant No. Y7292610K1.

\end{document}